\documentclass[12pt]{article}
\usepackage[dvips]{color}
\usepackage{epsfig}
\usepackage{amsmath}
\usepackage{graphicx}

\begin{document}

\title{Strong Gravitational Lensing in a Charged Squashed Kaluza- Klein Black hole}
\author{{J. Sadeghi $^{a}$\thanks{Email: pouriya@ipm.ir}\hspace{1mm}
{A. Banijamali $^{b}$\thanks{Email:
a.banijamali@nit.ac.ir}\hspace{1mm} and H. Vaez $^{a}$
\thanks{Email: h.vaez@umz.ac.ir}}}\\
{$^{a}$ \emph{Physics Department, Mazandaran University},}\\{
\emph{P.O.Box 47416-95447, Babolsar, Iran}}\\ {$^{b}$ \emph{
Department of Basic Sciences, Babol University of Technology, Babol,
Iran}}} \maketitle
\begin{abstract}
In this paper we investigate the strong gravitational lensing in a
charged squashed Kaluza-Klein black hole.  We suppose that the
supermassive black hole in the galaxy center can be considered by a
charged squashed Kaluza-Klein black hole and then we study the
strong gravitational lensing theory and estimate the numerical
values for parameters and observables of it. We explore the effects
of the scale of extra dimension $\rho_0$ and the charge of black
hole $\rho_q$ on these parameters and observables. \noindent
 \\
\end{abstract}
\section{Introduction}
We know  the light rays or photons would be deviated from their
straight way when they pass close to the massive object such as
black holes. This deflection of light rays well known as
gravitational lensing. This gravitational lensing is one of the
applications and results of general relativity \cite{Einstein} and
is used as an instrument in astrophysics because it can  help us to
extract the information about stars. The basics of gravitational
lensing theory developed by Liebes \cite{Liebes}, Refsdal
\cite{Refsdal}, Bourassa and Kantowski \cite{Bourassa}. The
gravitational lensing has been presented in details  in
\cite{schneider} and  reviewed  by some papers (see for examples
\cite{Narayan}-\cite{wabsganss}). At this stage, the gravitational
lensing is developed for weak field limit and could not describe
some phenomena such as looping of light rays near the massive
objects. Hence, scientists were starting to study these phenomena
from another point of view and they posed gravitational lensing in
strong field limit. Several studies about light rays close to the
Schwarzschild horizon have been done in literatures: a
semi-analytical investigation about geodesics in Kerr geometry has
been made in \cite{Viergutz}, also the appearance of a black hole in
front of a uniform background was studied in Refs.
\cite{Bardeen,Falcke}. Recently, Virbhadra and Ellis
\cite{Virbhadra} shown when a source be highly aligned with
Schwarzschild black hole and the observer, one set of infinitive
relativistic images would produce on each side of black hole. These
images are produced when the light ray passes with impact parameter
near to it,s minimum and winds one or several times around the black
hole before reaching to observer. In Refs \cite{Frittelli,bozza2},
the same problem was done  with other methods. Afterwards, the same
technique applied to other black holes \cite{Eiroa}-\cite{bozza1},
naked singularities and Janis-Newman-Winicour metric
\cite{bozza1,naked}.\\
Recently, the idea of large extra dimensions has attracted much
attention \cite{Arkani} to construct theories in which gravity be
unified with other forces. The five-dimensional Einstein-Maxwell
theory with a Chern- Simons term \cite{Gunaydin} predicted
five-dimensional charged black holes \cite{higherdimension}. Such a
higher-dimensional black holes would reside in a spacetime that is
approximately isotropic in the vicinity of the black holes, but
effectively four-dimensional far from the black holes
\cite{fourdim}. We call higher-dimensional black holes with this property Kaluza-Klein black holes.\\
Presence of extra dimension is tested by quasinormal modes from the
perturbation around the higher dimensional black hole \cite{QNM} and
the spectrum of Hawking radiation  \cite{Hawkingradiation}. The
gravitational lensing is another method to investigate the extra
dimension. Thus, the study of strong gravitational lensing by higher
dimensional black hole can help us to extract  information about
the extra dimension in astronomical observations in  the future. \\
The Kaluza-Klein black holes with squashed horizon is one of the
extra dimensional black holes and  it's Hawking radiation
 and quasinormal modes have been
investigated in some papers
\cite{HawkingradiationSKKBH1}-\cite{QNMSKKBH2}. Liu, Chen and Jing
have studied the gravitational lensing by squashed Kaluza-Klein
black holes  in Refs \cite{SKKBH,SKKGBH}. In Ref \cite{SKKBH}, the
Author probed the effect of extra dimension on parameters and
observables of strong gravitational lensing. Also
 variation of these parameters and observables  with G$\ddot{o}$del parameter and
 extra dimension is studied in a squashed Kaluza-Klein $G\ddot{o}$del black
hole in Ref \cite{SKKGBH}. In this paper we study the strong
gravitational lensing in a charged squashed Kaluza-Klein black hole
and peruse effects of the scale of extra dimension and charge of
black hole on the coefficients and observables of strong
gravitational lensing.\\
The rest of this paper is organized as follows: The Section 2 is
briefly devoted  to charged squashed Kaluza-Klein black hole metric.
In Section 3 we use the Bozza's method \cite{bozza5,bozza3} to
obtain the deflection angle and other parameters of strong
gravitational lensing as well as  variation of them with extra
dimension and charge of black hole . In Section 4, we suppose that
the supermassive object at the center of galaxy can be considered by
the metric of charged squashed Kaluza-Klein black hole. Then we
evaluate the numerical results for the coefficients and observables
in the strong gravitational lensing . In the last Section, we
present a summary of our work.
\section{The charged squashed Kaluza- Klein black hole metric}
The charged squashed Kaluza- Klein black hole spacetime is
considered as the metric \cite{Ishihara}
\begin{equation}\label{metric1}
ds^2=-f(r)dt^2+\frac{k^2(r)}{f(r)}dr^2+\frac{r^2}{4}[k(r)(\sigma^2_1+\sigma^2_2)+\sigma^2_3],
\end{equation}
where
\begin{eqnarray}\label{sigma}
&&\sigma_1=\cos\psi\, d\theta+\sin\psi\, \sin\theta\, d\phi,\nonumber\\
&&\sigma_2=-\sin\psi\,d\theta+\cos\psi\,\sin\theta\,d\phi,\nonumber\\
&&\sigma_3=d\psi+\cos\theta\, d\phi.
\end{eqnarray}
The coordinates run in range, $0\leq\theta<\pi$, $0\leq\phi<2\pi$,
$0\leq\psi<2\pi$ and $0<r<r_\infty$. Moreover, the functions
 in the metric define as
\begin{eqnarray}\label{function}
f(r)=1-\frac{2M}{r^2}+\frac{q^2}{r^4},\,\,\,\,\,\,\,
k(r)=\frac{f(r_\infty)r_\infty^4}{(r^2-r_\infty^2)^2}.
\end{eqnarray}
\begin{figure}[htp]
\begin{center}
\includegraphics{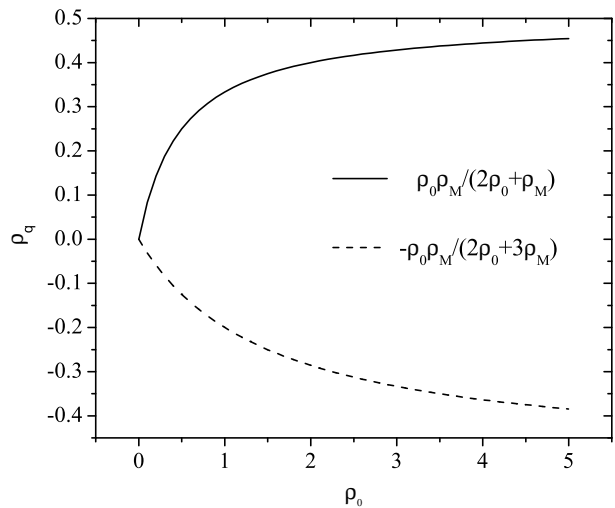} \vspace{7cm}\includegraphics{Hubble1.eps}
\end{center}
\caption{\scriptsize{The figure shows the admissible values for
$\rho_0$ and  $\rho_q$. Note that the region  between  two curves is
allowed.}}
\end{figure}
Here $M$ and $q$ are the mass and charge of the black hole
respectively . The killing horizon of the black hole is given by the
equation $f(r)=0$, that is $r_{h}^2=M\pm\sqrt{M^2-q^2}$. We see that
the black hole have two horizons. As $q\longrightarrow0$, one has
$r_{h}^2=2M$, which is the horizon of five-dimensional Schwarzschild
black hole, and for $M=q$ we have extremal black hole with a horizon
$r_{h\pm}^2=M$. Here we note that the argument of square root
constraints the mass and charge values  as, $|M|\geq|q|$. When
$r_{\infty}\longrightarrow \infty$, we have $k(r)\longrightarrow1$,
which means that the squashing effect
disappears and the five-dimensional charged black hole is recovered.\\
By using the transformations,
$\rho=\rho_0\frac{r^2}{r^2_\infty-r^2}$ and
$\tau=\sqrt{f(r_\infty)}t$, metric ~(\ref{metric1}) can be written
in the following form
\begin{equation}\label{metric2}
ds^2=-\mathcal{F}(\rho)d\tau^2+\frac{K(\rho)}{\mathcal{F}(\rho)}d\rho^2+\mathcal{C}(\rho)(d\theta^2+sin^2\theta\,d\phi^2)+\mathcal{D}(\rho)\sigma_3^2,
\end{equation}
\begin{eqnarray}\label{function2}
&\mathcal{F}(\rho)=(1-\frac{\rho_{h+}}{\rho})(1-\frac{\rho_{h-}}{\rho}),\nonumber\\
&K(\rho)=1+\frac{\rho_0}{\rho},\nonumber\\
&\mathcal{C}(\rho)=\rho^2 K(\rho),\,
\mathcal{D}(\rho)=\frac{r^2_\infty}{4K(\rho)},
\end{eqnarray}
\begin{figure}[htp]
\begin{center}
\includegraphics{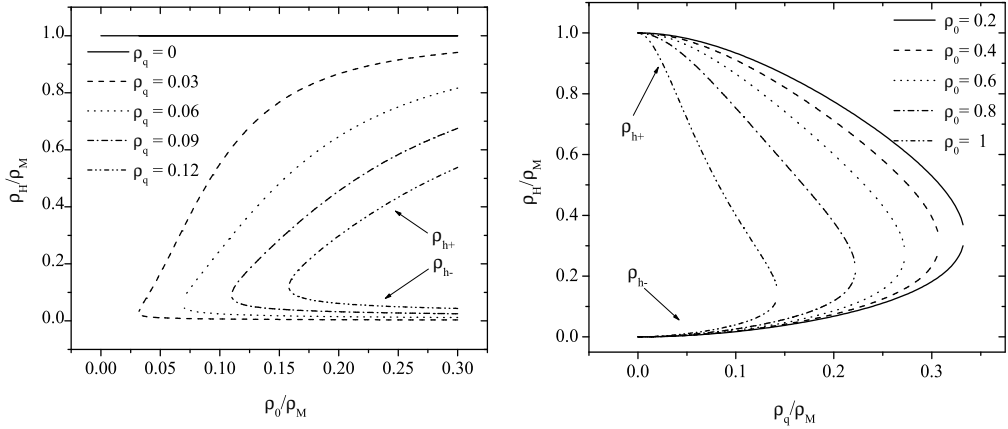} \vspace{7cm}\includegraphics{Hubble1.eps}
\end{center}
\caption{\scriptsize{The plots show
 variation  of radius of horizons with respect to $\rho_0$ and $\rho_q$.
}}
\end{figure}

where $\rho_{h+}$ and $\rho_{h-}$ denote the outer and inner
horizons of the black hole in the new coordinate and $\rho_0$ is a
scale of transition from five-dimensional spacetime to an effective
four-dimensional one. By using coordinate transformation we can
obtain $\rho_{h\pm}$ in the following form
\begin{equation}\label{horizon}
\rho_{h\pm}=\frac{2\rho_0}{2-\eta\mp\xi}-\rho_0
\end{equation}
where
\begin{eqnarray}\label{eta}
\eta=\frac{\rho_M}{\rho_0+\rho_M},\,\,\,\,\,\,\,
\xi=\sqrt{\frac{\rho_M^2}{(\rho_0+\rho_M)^2}-\frac{4\rho_q^2}{(\rho_0+\rho_q)^2}}\,\,\,\,,
\end{eqnarray}
with
\begin{eqnarray}\label{function2}
\rho_M=\rho_0\frac{2M}{r^2_\infty-2M}\,\,,
\,\,\,\,\,\,\,\rho_q=\rho_0\frac{q}{r^2_\infty-q}\,\,\,.
\end{eqnarray}
Note that
$\rho_{h\pm}=\rho_0\frac{r_{h\pm}^2}{r^2_\infty-r_{h\pm}^2}$. Here
$\rho_0^2=\frac{r^2_\infty}{4}V(r_\infty)$, so that
$r^2_\infty=4(\rho_0+\rho_{h+})(\rho_0+\rho_{h-})$. The Komar mass
of black hole is related to $\rho_M$ with $\rho_M=2G_4M$, where
$G_4$ is the four dimensional  gravitational constant. As
$\rho_q\longrightarrow0$, the horizon of black hole becomes
$\rho_h=\rho_M$, which is consistent with that in the neutral
squashed Kaluza-Klein black hole \cite{SKKBH}. We note that the
square root in relation ~(\ref{eta}) constrains the values of
$\rho_0$ and $\rho_q$. For
$\rho_q=-\frac{\rho_0\rho_M}{2\rho_0+3\rho_M}$ and
$\rho_q=\frac{\rho_0\rho_M}{2\rho_0+\rho_M}$, the extremal black
hole is obtained and admissible values for $\rho_q$ are between
these two values. We have plotted the admissible region for $\rho_0$
and $\rho_q$  in figure 1. Also we have evaluated the admissible
values for specific values, $\rho_0$ and $\rho_q$, in table 1.
Hereafter, our figures are plotted  by considering the table 1.
 Variation of the radius of
horizons are plotted in figure 2.
\begin{table*}\label{s}
\begin{center}
\begin{tabular}{ c c c c c | c c c c c}
  \hline
  \hline
    $\rho_q$      &&    $ \rho_0$      & &   {} & {} &&    $ \rho_0$     &&       $\rho_q$          \\
  \hline
      $0$   &&$\rho_0>0    $   &&&&&     0    &&     $\rho_q=0$  \\
   $0.03$   &&$\rho_0>0.032$   &&&&&    0.2   &&     $-0.059<\rho_q<0.143$   \\
   $0.06$   &&$\rho_0>0.068$   &&&&&    0.4   &&     $-0.105<\rho_q<0.222$  \\
   $0.09$   &&$\rho_0>0.110$   &&&&&    0.6   &&     $-0.143<\rho_q<0.273$  \\
   $0.12$   &&$\rho_0>0.158$   &&&&&    0.8   &&     $-0.174<\rho_q<0.308$       \\
   $0.15$   &&$\rho_0>0.214$   &&&&&     1    &&     $-0.200<\rho_q<0.333$    \\
  \hline
\end{tabular}
\caption {Admissible values for $\rho_0$ and $\rho_q$ for constants
values of $\rho_q$ and $\rho_0$ respectively.}
\end{center}
\end{table*}

\section{Deflection angle in the strong field limit}
In this section, we will investigate deflection angle of the light
rays when they pass close to a charged squashed Kaluza-Klein black
hole and probe the effect of the charge parameter $\rho_q$ and the
scale of extra dimension $\rho_0$ on the deflection angle and
coefficients of it in the equatorial plan
$\theta=\pi/2$.\\
We consider the metric for charged squashed Kaluza-Klein black hole
as
\begin{equation}\label{metric3}
ds^2=-\mathcal{F}(\rho)d\tau^2+\mathcal{B}(\rho)d\rho^2+\mathcal{C}(\rho)\,d\phi^2+\mathcal{D}(\rho)d\psi^2,
\end{equation}
where
\begin{equation}\label{B}
\mathcal{B}(\rho)=\frac{K(\rho)}{\mathcal{F}(\rho)}.
\end{equation}
The null geodesics equations are
\begin{equation}\label{geodesic}
\frac{d\upsilon_i}{dk}+\Gamma_{jk}^i\,\upsilon^j\,\upsilon^k=0,
\end{equation}
where
\begin{equation}\label{geodesic2}
g_{ij}\upsilon^i\,\upsilon^j=0.
\end{equation}
$\upsilon^i=\frac{dx^i}{dk}$ is the tangent vector to the null
geodesics and $k$ is affine parameter. The following equations can
be obtained from~(\ref{geodesic}),
\begin{eqnarray}\label{constMo}
\frac{dt}{dk}=\frac{E}{\mathcal{F}(\rho)},\nonumber\\
\frac{d\phi}{dk}=\frac{L_\phi}{\mathcal{C}(\rho)},\nonumber\\
\frac{d\psi}{dk}=\frac{L_\psi}{\mathcal{D}(\rho)},
\end{eqnarray}
\begin{equation}\label{rhoGeo}
(\frac{d\rho}{dk})^2=\frac{1}{\mathcal{B}(\rho)}\left[\frac{\mathcal{D}(\rho)E-\mathcal{F}(\rho)L^2_\psi}{\mathcal{F}(\rho)\mathcal{D}(\rho)}-\frac{L^2_\phi}{\mathcal{C}(\rho)}\right].
\end{equation}
Here $E$, $L_\phi$ and $L_\psi$ are constants of motion. Also, the
$\theta$-component of equation~(\ref{geodesic}) in the equatorial
plan, $\theta=\pi/2$, leads to
\begin{equation}\label{s14}
\frac{d\phi}{dk}(\mathcal{D}(\rho)\frac{d\psi}{dk})=0.
\end{equation}
So this constraint implies that either $\frac{d\phi}{dk}=0$ or
$L_\psi=\mathcal{D}(\rho)\frac{d\psi}{dk}=0$. Now, we set
$L_\psi=0$, as
done in Refs.  \cite{SKKGBH,SKKBH}.\\
From equation~(\ref{rhoGeo}), the impact parameter and photon sphere
equation can be obtained as
\begin{equation}\label{impact}
u=\sqrt{\frac{\mathcal{C}(\rho)}{\mathcal{F}(\rho)}},
\end{equation}
and
\begin{equation}\label{phEqu}
\mathcal{F}(\rho)\,\mathcal{C}^\prime(\rho)-\mathcal{C}(\rho)\,\mathcal{F}^\prime(\rho)=0.
\end{equation}
\begin{figure}[htp]
\begin{center}
\includegraphics{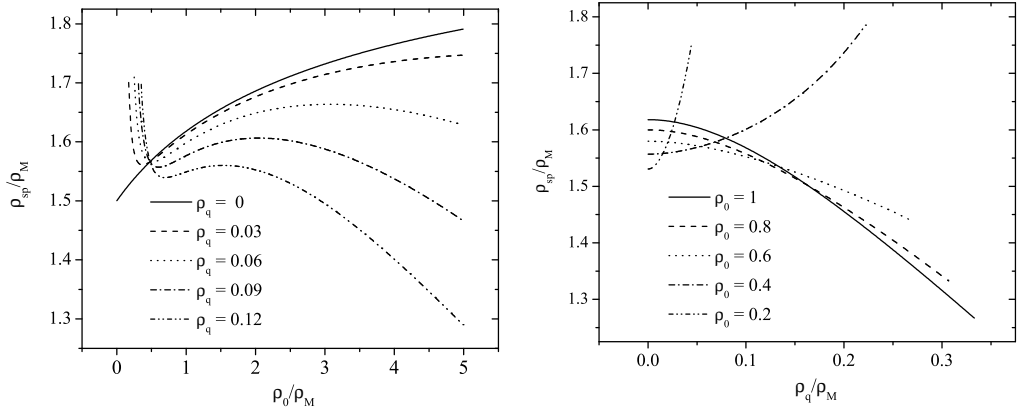} \vspace{7cm}\includegraphics{Hubble1.eps}
\end{center}
\caption{\scriptsize{The figures show the variation  of radius of
photon sphere with respect to $\rho_0$ and $\rho_q$. }}
\end{figure}
\begin{figure}[htp]
\begin{center}
\includegraphics{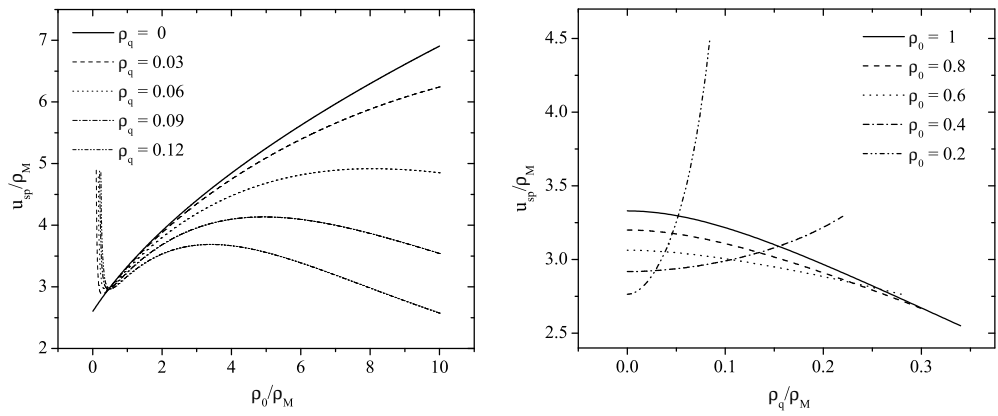} \vspace{7cm}\includegraphics{Hubble1.eps}
\end{center}
\caption{\scriptsize{Variations of impact parameter in the charged
squashed Kaluza-Klein black hole.}}
\end{figure}

The above equation is very complicated to solve analytically  and
thus we have calculated it numerically. Variations of radius of
photon sphere  are plotted  with respect to the charge $\rho_q$ and
the scale of extra dimension $\rho_0$ in the figure 3. Figure 4
shows variation of impact parameter in it's minimum value (at radius
of photon sphere) with respect to the same parameters . These
figures show that by adding the charge to the black hole the
behavior of radius of photon sphere and minimum of impact parameter
is different in compare with the neutral black hole \cite{SKKBH}. As
$\rho_0$ approaches to it's minimum values in the table 1, the
radius of photon sphere and impact parameter become divergent and
when $\rho_0$ increases, $\rho_{sp}$ and $u_{sp}$ increase up to a
maximum and descend. Also, when $\rho_q$   increases then,
$\rho_{sp}$ and $u_{sp}$ increase for smaller $\rho_0$ and decrease
for the larger $\rho_0$. Variation of the impact parameter and
radius of photon sphere are similar, but note that
 for $\rho_q=0$,  when
$\rho_0\longrightarrow\infty$, the photon sphere curve tends to
$\rho_{sp}=2$, while the impact parameter goes to infinity.\\
The deflection angle in the charged squashed Kaluza-Klein black hole
can be written  as
\begin{equation}\label{s17}
\alpha(\rho_s)=I(\rho_s)-\pi,
\end{equation}
where $\rho_s$ is the closest approach and
\begin{eqnarray}\label{I}
I(\alpha)=2\int^\infty_{\rho_s}\left[\frac{\mathcal{C}(\rho)}{\mathcal{C}(\rho_s)}\mathcal{F}(\rho_s)-\mathcal{F}(\rho)\right]^{-\frac{1}{2}}\frac{d\rho}{\rho}.
\end{eqnarray}
When we decrease the $\rho_s$ (and consequently $u$) the deflection
angle increases. At some points, the deflection angle will exceed
from $2\pi$ so that the light ray makes a complete loop around the
compact object before reaching the observer. By decreasing $\rho_s$
further, the photon will wind several times before emerging.
Finally, for $\rho_s=\rho_{sp}$ the deflection angle diverges and
the photon is
captured by the black hole.\\
   We can rewrite the equation~(\ref{I}) as

\begin{equation}\label{Iz}
I(\rho_s)=2\int^1_0R(z,\rho_s)f(z,\rho_s)\,dz,
\end{equation}
with
\begin{equation}\label{R}
R(z,\rho_s)=2\frac{\rho}{\rho_{s}}\sqrt{\frac{\mathcal{C}(\rho_s)}{\mathcal{C}(\rho)}},
\end{equation}
and
\begin{equation}\label{f(z)}
f(z,\rho_s)=\frac{1}{\sqrt{\mathcal{F}(\rho_s)-\mathcal{F}\frac{\mathcal{C}(\rho_s)}{\mathcal{C}(\rho)}}},
\end{equation}
where we have defined $z=1-\frac{\rho_s}{\rho}$. The function
$R(z,\rho_s)$ is regular for all values of $z$ and $\rho_s$, while
$f(z,\rho_s)$ diverges as $z$ approaches to zero. Therefore, we can
split the integral~(\ref{I}) in two part, the divergent part
$I_D(\rho_s)$ and the regular one $I_R(\rho_s)$, as follows
\begin{equation}\label{Id}
I_D(\rho_s)=\int^1_0R(0,\rho_{sp})f_0(z,\rho_s)\,dz,
\end{equation}
\begin{equation}\label{Ir}
I_R(\rho_s)=\int^1_0\left[R(z,\rho_s)f(z,\rho_s)-R(0,\rho_{sp})f_0(z,\rho_s)\right]\,dz.
\end{equation}
Here we expand the argument of the square root in $f(z,\rho_s)$ up
to the second order in $z$ \cite{SKKGBH}
\begin{equation}\label{f0}
f_0(z,\rho_s)=\frac{1}{\sqrt{p(\rho_s)z+q(\rho_s)z^2}},
\end{equation}
where
\begin{equation}\label{p}
p(\rho_s)=\frac{\rho_s}{\mathcal{C}(\rho_s)}\left[\mathcal{C}^\prime(\rho_s)\mathcal{F}(\rho_s)-\mathcal{C}(\rho_s)\mathcal{F}^\prime(\rho_s)\right],
\end{equation}
\begin{equation}\label{q}
q(\rho_s)=\frac{\rho_s^2}{2\mathcal{C}(\rho_s)}\left[2\mathcal{C}^\prime(\rho_s)\mathcal{C}(\rho_s)\mathcal{F}^\prime(\rho_s)
-2\mathcal{C}^\prime(\rho_s)^2\mathcal{F}(\rho_s)+\mathcal{F}(\rho_s)\mathcal{C}(\rho_s)\mathcal{C}^{\prime\prime}(\rho_s)-
\mathcal{C}^2(\rho_s)\mathcal{F}^{\prime\prime}(\rho_s)\right].
\end{equation}
For $\rho_s>\rho_{sp}$, $p(\rho_s)$ is nonzero and the leading order
of the divergence in $f_0$ is $z^{-1/2}$, which have a finite
result. As $\rho_s\longrightarrow\rho_{sp}$, $p(\rho_s)$ approaches
zero and divergence is of order $z^{-1}$, that makes the integral
divergent. Therefor, the deflection angle can be approximated in the
following form \cite{bozza1}
\begin{equation}\label{deflection}
\alpha(\theta)=-\bar{a}\,log\left(\frac{\theta
D_{OL}}{u_{sp}}-1\right)+\bar{b}+O(u-u_{sp}),
\end{equation}
where
\begin{eqnarray}\label{a}
&&\bar{a}=\frac{R(0,\rho_{sp})}{2\sqrt{q(\rho_{sp})}}\,,\nonumber\\
&&\bar{b}=-\pi+b_R+\bar{a}\,log\frac{\rho_{sp}^2\left[\mathcal{C}^{\prime\prime}(\rho_{sp})\mathcal{F}(\rho_{sp})-
\mathcal{C}(\rho_{sp})\mathcal{F}^{\prime\prime}(\rho_{sp})\right]}{u_{sp}\sqrt{\mathcal{F}^3(\rho_{sp})\mathcal{C}(\rho_{sp})}}\,,\nonumber\\
&&b_R=I_R(\rho_{sp}),\,\,\,\,\,\,u_{sp}=\sqrt{\frac{\mathcal{C}(\rho_{sp})}{\mathcal{F}(\rho_{sp})}}\,.
\end{eqnarray}
The parameter $D_{OL}$ is the distance between an observer and
gravitational lens. Using of~(\ref{deflection}) and~(\ref{a}), we
can investigate the properties of strong gravitational lensing in
the charged squashed Kaluza- Klein black hole. In this case,
variations of the $u_{sp}$, the coefficients $\bar{a}$ and
$\bar{b}$, and the deflection angle $\alpha(\theta)$ have been
plotted with respect to
 the extra dimension $\rho_0$, and charge of the black hole $\rho_q$, in
 figures 5-7.\\
 In figures 5 and 6, we see that for fixed $\rho_q$, the coefficient $\bar{a}$ increases with
 the size of the extra dimension and the coefficient $\bar{b}$ increases for the
 smaller $\rho_q$ and decreases for the larger ones. Moreover, $\bar{a}$ decreases with
 increasing $\rho_q$ for fixed $\rho_0$. The coefficient $\bar{b}$ increases up to a maximum and descends
 for smaller $\rho_0$ but increases monotonically for larger $\rho_0$.
 Variation of deflection angle is presented in figure 7.
 One can see that the deflection angle increases with extra
 dimension  and decreases with $\rho_q$. We note that, as $\rho_0$ approaches to it's minimum values in left column of table 1,
 the coefficients $\bar{a}$, $\bar{b}$ and deflection angle $\alpha(\theta)$ tend to zero. By comparing these parameters with
 those in four-dimensional schwarzschild and  Reissner-Nordstr$\ddot{o}$m black
 holes
 , we could extract information about the size of
 extra dimension as well as the charge of black hole by using strong field
 gravitational lensing.
 \begin{figure}[htp]
\begin{center}
\includegraphics{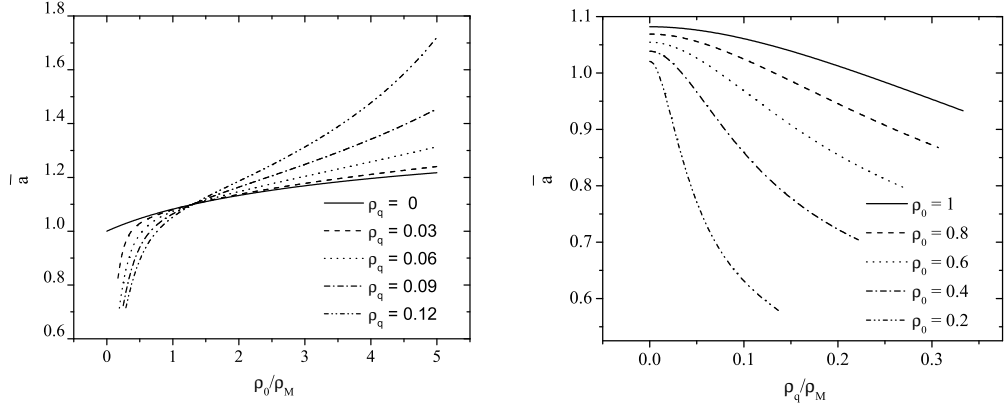} \vspace{7cm}\includegraphics{Hubble1.eps}
\end{center}
\caption{\scriptsize{Variation of the coefficient $\bar{a}$ with
respect to the scale parameter $\rho_0$ and charge of black hole
$\rho_q$. }}
\end{figure}

\begin{figure}[htp]
\begin{center}
\includegraphics{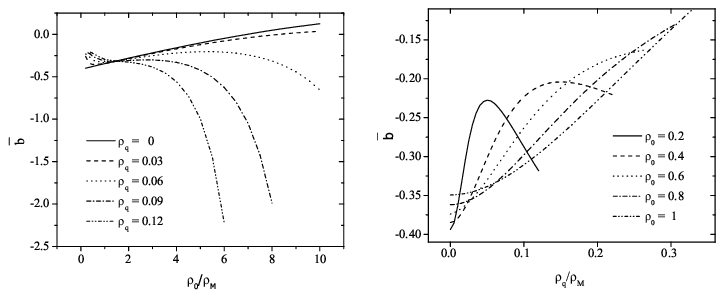} \vspace{7cm}\includegraphics{Hubble1.eps}
\end{center}
\caption{\scriptsize{Variation of the coefficient $\bar{b}$ with
respect to the scale parameter $\rho_0$ and charge of black hole
$\rho_q$. }}
\end{figure}

\begin{figure}[htp]
\begin{center}
\includegraphics{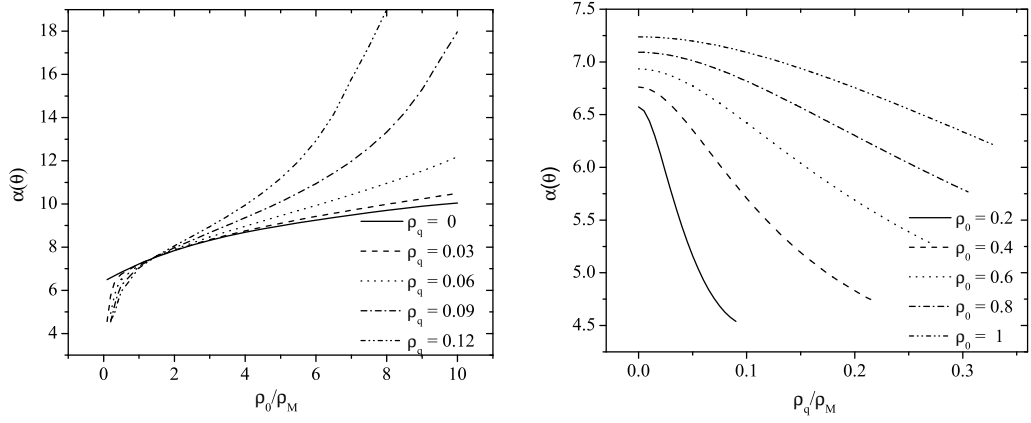} \vspace{7cm}\includegraphics{Hubble1.eps}
\end{center}
\caption{\scriptsize{The plots show the variation of deflection
angle with $\rho_0$ and $\rho_q$ in the charged squashed
Kaluza-Klein black hole. ( we considered $u=u_{sp}+0.003$ )}}
\end{figure}

\begin{table*}[s]
\begin{center}
\begin{tabular}{c | c | c c c c c}
  \hline
  \hline
    $\rho_q$      &    $\rho_0$       &      $\theta_{\infty}$ \,($\mu$ arcsec)   &     $s$\,($\mu$ arcsec)       &   $r_m$ \,(magnitudes)  & $\bar{a}$  &  $\bar{b}$  \\

  \hline
  {Schwarzschild$\longrightarrow$}   &  $0$   &16.869  &  0.02111    &  6.8218 & 1.0 & -0.4003   \\
  {}   &  $0.2$   &17.949  &  0.02588    &  6.6838 & 1.0206 & -0.3938   \\
  {}   & $0.4$    &18.950 & 0.03088 & 6.5678 & 1.0387 & -0.3847   \\
  0    & $0.6$    &19.890 & 0.03610 & 6.4681 & 1.0547 & -0.3738  \\
  {}   & $0.8$    &20.778 & 0.04150 & 6.3813 & 1.0690 & -0.3619   \\
  {}   & $1$      &21.623 & 0.04708 & 6.3046 & 1.0820 & -0.3493   \\
  \hline
  {}   &  $0.2$    & 19.098 & 0.01091 & 7.787 & 0.8760 & -0.2590    \\
  {}   & $0.4$     &18.998 & 0.02638 & 6.7706 & 1.0076 & -0.3460  \\
  0.03 & $0.6$     &19.848 & 0.03416 & 6.5376 & 1.0435 & -0.3585   \\
  {}   & $0.8$     &20.714 & 0.04047 & 6.4113 & 1.0640 & -0.3543 \\
  {}   & $1$       &21.546 & 0.04651 & 6.3174 & 1.0799 & -0.3452   \\
  \hline
  {}   &  $0.2$   & 22.698 & 0.00307 & 9.3261 & 0.7315 & -0.2318    \\
  {}   & $0.4$    &  19.127 & 0.01828 & 7.2268 & 0.9440 & -0.2805    \\
  0.06 & $0.6$    & 19.737 & 0.02691& 6.7145 & 1.0160 & -0.3231     \\
  {}   & $0.8$    & 20.541 & 0.03780 & 6.4921 & 1.0508 & -0.3346     \\
  {}   & $1$      &21.333 & 0.04495 & 6.3531 & 1.0738 & -0.3340   \\
  \hline
  {}   & $0.2$ &  31.798 & 0.00133 & 10.4912 & 0.6502 & -0.2718   \\
  {}   & $0.4$ &  19.332 & 0.01164 & 7.7627 & 0.8788 & -0.2333   \\
  0.09 & $0.6$ &  19.576 & 0.02422 & 6.9563 & 0.9807 & -0.2825    \\
  {}   & $0.8$ &  20.284 & 0.03415 & 6.6107 & 1.0319 & -0.3076   \\
  {}   & $1$ &    21.011 & 0.04263 & 6.4082 & 1.0646 & -0.3172     \\
  \hline
  {}   & $0.2$ &   72.561 & 0.00121 & 11.3647 & 0.6003 & -0.3184    \\
  {}   & $0.4$ &   19.619 & 0.00736 & 8.2875 & 0.8232 & -0.2101   \\
  0.12 & $0.6$ &   19.379 & 0.01915 & 7.2306 & 0.9435 & -0.2455  \\
  {}   & $0.8$ &   19.662 & 0.03008 & 6.7562 & 1.0097 & -0.2776   \\
  {}   & $1$ &     20.606 & 0.03980 & 6.4801 & 1.0527 & -0.2961 \\
  \hline
\end{tabular}
\caption {Numerical estimations for the coefficients and observables
of strong gravitational lensing by considering the supermmasive
object of galactic center be a charged squashed Kaluza-Klein black
hole.}
\end{center}
\end{table*}
%*****************************************************************
\section{Observables in the strong field limit }
Now, we are going to study the effects of the scale parameter
$\rho_0$ and charge of black hole $\rho_q$ on the observables in the
strong gravitational lensing. If we suppose that the spacetime of
the supersessive black hole at the galaxy center of Milky Way can be
considered by a charged squashed Kaluza-Klein black hole then, we
can estimate the numerical values for the coefficients and
observables of gravitational lensing in the strong field limit.\\
We can write the lens equation in strong gravitational lensing, as
the source, lens, and observer are highly aligned as follows
\cite{bozza2}
\begin{equation}\label{lensEQ}
\beta=\theta-\frac{D_{LS}}{D_{OS}}\Delta\alpha_n,
\end{equation}
\begin{figure}[htp]
\begin{center}
\includegraphics{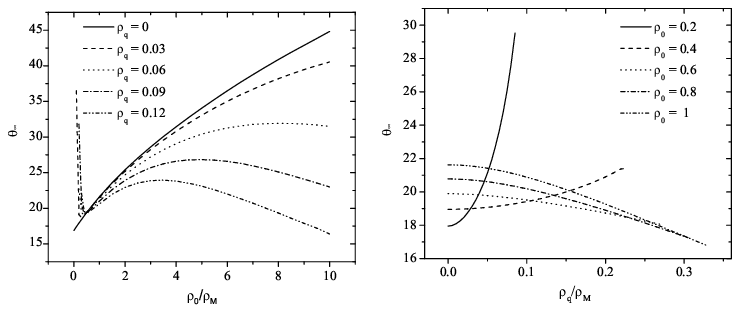} \vspace{7cm}\includegraphics{Hubble1.eps}
\end{center}
\caption{\scriptsize{Variation of the angular position
$\theta_{\infty}$ with the scale parameter $\rho_0$ and charge of
black hole $\rho_q$. }}
\end{figure}

\begin{figure}[htp]
\begin{center}
\includegraphics{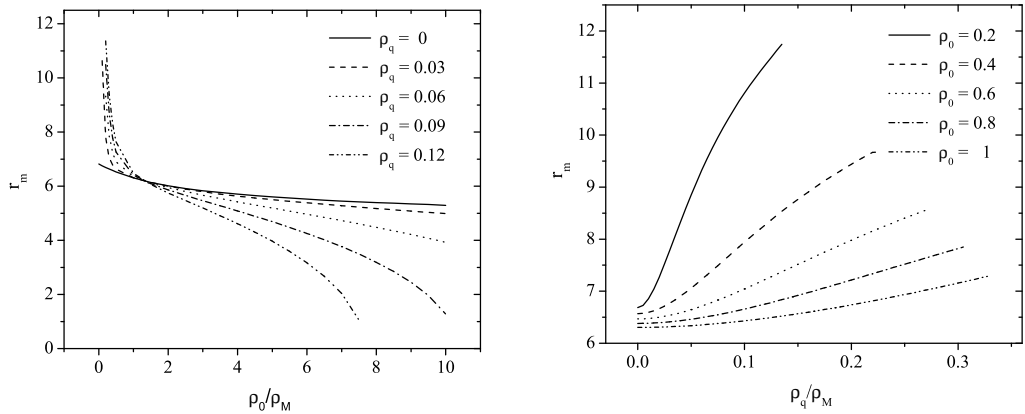} \vspace{7cm}\includegraphics{Hubble1.eps}
\end{center}
\caption{\scriptsize{Variation of the relative magnitudes  $r_{m}$
with the scale parameter $\rho_0$ and charge of black hole $\rho_q$.
}}
\end{figure}

\begin{figure}[htp]
\begin{center}
\includegraphics{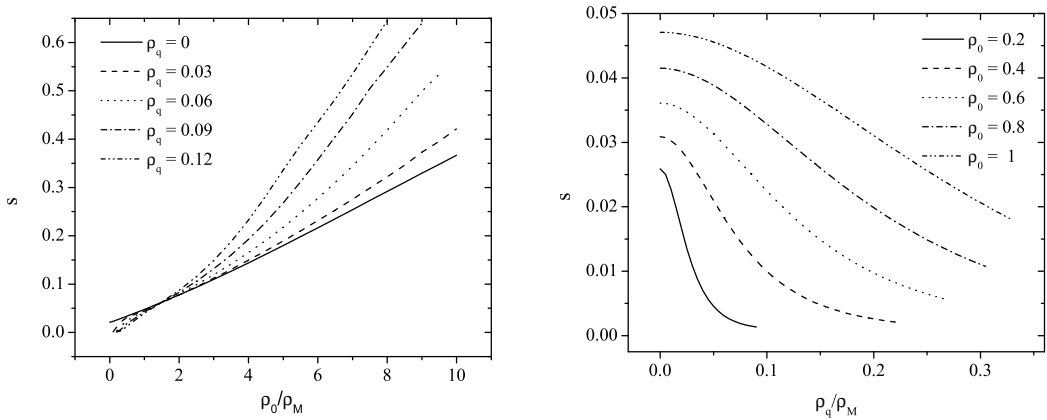} \vspace{7cm}\includegraphics{Hubble1.eps}
\end{center}
\caption{\scriptsize{Variation of the angular separation  $s$ with
the scale parameter $\rho_0$ and charge of black hole $\rho_q$. }}
\end{figure}
where $D_{LS}$ is the distance between the lens and source. $D_{OS}$
is the distance between the observer and the source so that,
$D_{OS}=D_{LS}+D_{OL}$. $\beta$ and $\theta$ are the angular
position of the source and the image with respect to lens,
respectively. $\Delta\alpha_n=\alpha-2n\pi$ is the offset of
deflection angle
with integer $n$  which indicate the $n$-th image.\\
The $n$-th image position $\theta_n$ and the $n$-th image
magnification $\mu_n$ can be approximated as follows
\cite{bozza1,bozza2}
\begin{equation}\label{theta}
\theta_n=\theta^0_n+\frac{u_{sp}(\beta-\theta_n^0)e^{\frac{\bar{b}-2n\pi}{\bar{a}}}D_{OS}}{\bar{a}
D_{LS}D_{OL}},
\end{equation}
\begin{equation}\label{magnification}
\mu_n=\frac{u_{sp}^2(1+e^{\frac{\bar{b}-2n\pi}{\bar{a}}})e^{\frac{\bar{b}-2n\pi}{\bar{a}}}D_{OS}}{\bar{a}\beta
D_{LS}D_{OL}^2}.
\end{equation}
$\theta_n^0$ is the angular position of $\alpha=2n\pi$. In the limit
$n\longrightarrow\infty$, the relation between the minimum of impact
parameter $u_{sp}$ and asymptotic position of a set of images
$\theta_{\infty}$ can be expressed by
$u_{sp}=D_{OL}\theta_{\infty}$. In order to obtain the coefficients
$\bar{a}$ and $\bar{b}$, in the simplest case, we separate the
outermost image $\theta_1$ and all the remaining ones which packed
together at $\theta_{\infty}$, as done in Refs \cite{bozza1,bozza2}.
Thus $s=\theta_1-\theta_{\infty}$ is considered as the angular
separation between the first image and other ones and the ratio of
the flux of them is given by
\begin{equation}\label{R}
\mathcal{R}=\frac{\mu_1}{\sum_{n=2}^{\infty}\mu_n}.
\end{equation}
We can simplify the observables and rewrite them in the following
form
\begin{eqnarray}\label{sR}
&&s=\theta_{\infty}e^{\frac{\bar{b}}{\bar{a}}-\frac{2\pi}{\bar{a}}},\nonumber\\
&&\mathcal{R}=e^{\frac{2\pi}{\bar{a}}}.
\end{eqnarray}
Thus, by measuring the $s$, $\mathcal{R}$ and $\theta_{\infty}$, one
can obtain the  values of the coefficients $\bar{a}$, $\bar{b}$ and
$u_{sp}$. If we compare these  values by those obtained  in the
previous section, we could detect the size of the extra dimension
and charge of black hole.  Another  observable for gravitational
lensing is relative magnification of the outermost relativistic
image with the other ones. This observable is showed by $r_m$ which
is related to $\mathcal{R} $ by
\begin{eqnarray}\label{rm}
r_m=2.5\, \log\mathcal{R}.
\end{eqnarray}
Using $\theta_{\infty}=\frac{u_{sp}}{D_{OL}}$ and
equations~(\ref{a}),~(\ref{sR}) and~(\ref{rm}) we can estimate the
values of the coefficients $\bar{a}$ and $\bar{b}$, in the strong
field gravitational lensing. Variation of  the observables
$\theta_{\infty}$, $s$ and $r_m$ are plotted in figures 8-10. Note
that the mass of the central object of our galaxy is estimated to be
$2.8\times 10^6 M_\odot$ and  the distance between the sun and the
center of galaxy is $D_{OL}=8.5\,kpc$ \cite{Virbhadra}.\\
For the different $\rho_0$ and $\rho_q$, the numerical values for
main observables and the strong field limit coefficients for the
black hole at center of our galaxy which is supposed to be described
by the charged squashed Kaluza-Klein black hole are listed in Table
2. One can see that our results reduce to those in the
four-dimensional Schwarzschild black hole as $\rho_0=0$ and also our
results are in agreement with the results of Ref. \cite{SKKBH} in
the limit $\rho_q\longrightarrow0$.
\section{Summary}
The extra dimension in one of the important predictions in string
theory which is believed to be promising candidate for the unified
theory. The five-dimensional Einstein-Maxwell theory with a Chern-
Simons term in string theory  predicted five-dimensional charged
black holes. We considered the charged squashed Kaluza-Klein black
hole spacetime and investigated the strong gravitational lensing by
this metric. We obtained theoretically the deflection angle and
other parameters of strong gravitational lensing and studied
variation of them with respect to extra dimension and charge of
black hole. Finally we estimated numerically the values of
observables i.e. relativistic images $\theta_{\infty}$, the angular
separation $s$ and the relative magnitudes $r_m$ for different
$\rho_0$ and $\rho_q$. Our results are presented in figures 1-10 and
Table 2. These results may help us to detect the extra dimension and
charge of black hole by astronomical observations in  the future.
\subsection*{Acknowledgments}
 H. V would like to thank Reza Gharai for helpful comments on our numerical calculations.

\end{document}